# Analytical model for large-scale design of sidewalk delivery robot systems

Hai Yang[1], Yuchen Du[2], Tho V. Le[2], Joseph Y. J. Chow*[1]

[1]C2SMARTER University Transportation Center, Department of Civil and Urban Engineering, New York University Tandon School of Engineering, Brooklyn NY, USA

[2]School of Engineering Technology, Purdue University, West Lafayette, IN, USA

*Corresponding author email: joseph.chow@nyu.edu

**Abstract**

With the rise in demand for local deliveries and e-commerce, robotic deliveries are being considered as efficient and sustainable solutions. However, the deployment of such systems can be highly complex due to numerous factors involving stochastic demand, stochastic charging and maintenance needs, complex routing, etc. We propose a model that uses continuous approximation methods for evaluating service trade-offs that consider the unique characteristics of large-scale sidewalk delivery robot systems used to serve online food deliveries. The model captures both the initial cost and the operation cost of the delivery system and evaluates the impact of constraints and operation strategies on the deployment. By minimizing the system cost, variables related to the system design can be determined. First, the minimization problem is formulated based on a homogeneous area, and the optimal system cost can be derived as a closed-form expression. By evaluating the expression, relationships between variables and the system cost can be directly obtained. We then apply the model in neighborhoods in New York City to evaluate the cost of deploying the sidewalk delivery robot system in a real-world scenario. The results shed light on the potential of deploying such a system in the future.

**ACKNOWLEDGEMENTS**

The authors wish to acknowledge the funding support from C2SMARTER University Transportation Center (USDOT #69A3551747124) for the first and last authors.

# 1. Introduction

The rise of the online food delivery (OFD) industry is evident: revenue gathered in the global market is projected to be 1.02 trillion US dollars in 2023, which is nearly 4 times the number in 2017 (Statista, 2023). In the United States (US) market, the revenue of the OFD industry also increased by 260% in the past 5 years (Statista, 2023). By 2027, the US market revenue is projected to increase by another 75%. At the same time, the penetration rate could reach 55.0%, up from 43.9% in 2022 (Statista, 2023).

The rapidly expanding market of the OFD industry poses a significant challenge to last-mile delivery operations. Currently, such big platforms as DoorDash and Uber Eats dominate OFD delivery (MCkinsey & Company, 2023), and they rely on on-demand workers to offer last-mile service to customers. The gig economy has persistently been put under the spotlight when discussing the social impact of the OFD industry. Food delivery workers consistently face low payment and unfair treatment from both the platform company and the customers (Milkman et al., 2021). For the big platform companies, higher market share can be achieved by offering faster service to customers, which helps generate more demand. However, high time constraints are often placed on food delivery workers to fulfill the fast service commitment, which could significantly increase unsafe on-road practices (Lin and Jia, 2023). From the customer's aspect, the additional fee for food delivery services induces an additional 10% or more extra spending per order (MCkinsey & Company, 2023), limiting its expansion to lower income groups.

Emerging technologies offer new alternatives for providing OFD services, and a sidewalk delivery robot (SDR) system is one of the highly discussed options. With the development of autonomous driving technologies, box-size robots navigating through sidewalks for food and grocery deliveries start to become a common scene in real-world applications. Starship, one of the pioneers in providing OFD using SDRs, provides local-level services in Europe and the US (Starship, 2023). Serve Robotics announced that it is launching a partnership with Uber to provide robotic food delivery services in Los Angeles (Serve, 2023). By deploying these robots, more standardized service experiences can be provided with better control of the fleets. Potential savings from eliminating labor costs can also be beneficial to both operators and customers.

At the same time, local legislators are introducing regulations regarding SDR operations. In the US, at least 30 states have either proposed or passed SDR-dedicated regulations (Clamann, 2023). Restrictions including speed limit and robot dimensions are commonly addressed by the bills. In addition, SDR's functionalities and operation capabilities are also defined in the regulations. In such way, dangerous on-road behaviors could be avoided, providing a better traffic environment to both pedestrians and vehicles.

Currently, all SDR deployments are in pilot phases, and companies are still exploring ways to efficiently operate SDR fleets. The literature that evaluates the applications of SDR in OFD is also lacking, preventing us from gaining more understanding of the potential of mass deployment. The objective of this study is to capture the average system costs of installing and operating a large-scale SDR fleet to provide OFD services. A continuous approximation (CA) model is developed to fulfill the purpose. The model can help determine the optimal SDR fleet size given predetermined service demands, service regions, and depot locations. By using this model, the impact of different factors like level of demand, pedestrian densities, operation strategies, and robot capabilities, can be quantified in terms of total system cost. Evaluation of potential SDR deployment in New York City (NYC) neighborhoods using synthetic data is conducted based on the model. The result provides practical insights into deploying an SDR system in the OFD industry, providing valuable information to both policymakers and operators on its potential.

The remainder of this paper is organized as follows. In the following section, we present a literature review of related studies. Next, we develop a model that can properly estimate the system cost of a proposed sidewalk robot online food delivery (SROFD) service. This model enables us to derive an optimal system cost, which is then formulated in a closed form. In the subsequent section, the formulation is investigated to understand the relationship between cost and different variables. A case study is conducted based on a real-world scenario, and the results are compared with the service fee charged by the existing OFD platforms. Finally, we conclude the study and discuss potential future directions.

## 2. Literature Review

### 2.1 Online food delivery

The meal delivery routing problem (MDRP) is a category of routing problems associated with the OFD industry, which has gathered increasing attention from the academic field. It is often categorized as a dynamic vehicle routing problem (DVRP), which involves formulating dynamic pickup and delivery operations. The delivery route needs to involve proper sequencing of pickup and delivery stops, making the problem more complex than the original vehicle routing problem (VRP). Reyes et al. (2018) introduced an optimization algorithm that addresses the courier assignment and capacity management problems using a rolling horizon strategy. Ulmer, Thomas, and Mattfeld (2019) developed a dynamic delivery strategy that allows preemptive depot returns before delivering all onboard orders.

Yildiz and Savelsbergh (2019) introduced a formulation for the MDRP with a solution algorithm by using simultaneous column and row generation methods. However, the formulation relies on perfect information about the order arrivals. To study the MDRP under uncertain demands, Steever et al. (2019) formulated a mixed integer linear programming to define an MDRP. A heuristic algorithm was proposed to solve the formulation in a dynamic setting, which proactively seeks solutions by anticipating future system states. Ulmer et al. (2021) developed a route-based Markov decision problem to model an MDRP dynamic uncertainty. By adding controlled postponement decisions, its solution method finds dispatch and routing strategies that meet the order deadline while addressing the problem of uncertain order placement time and food ready time.

Due to the complexity of the problems, the size of the experiments involved in the studies is generally small. In Yildiz and Savelsbergh (2019), the largest test sample involves 505 orders and 116 restaurants. In Steever et al. (2019) and Ulmer et al. (2021), the total number of orders in the experiment are both less than 300. To apply these methodologies in a large-scale system involving thousands of orders, excessive computational power and model-solving time are required.

### 2.2 Ground autonomous delivery

With the advancement of autonomous technologies, the application of using either autonomous vehicles or smaller ground autonomous robots to conduct last-mile services has gained more attention. Liu et al. (2020) introduced autonomous delivery vehicles (ADVs) into the E-grocery distribution system to reduce transport costs and fuel emissions. The objective of the study is to optimize a two-echelon distribution network for efficient E-grocery delivery, in which traditional vans serve the delivery in the first echelon and ADVs serve in the second echelon. The researchers formulated this problem as a two-echelon vehicle routing problem with mixed vehicles (2E-VRP-MV) with a nonlinear objective function, which optimizes the total transport and emission costs, and proposed a two-step clustering-based hybrid Genetic Algorithm and Particle Swarm Optimization (C-GA-PSO) algorithm to solve this problem. They observed that lower costs are achieved when customer density is lower and the depot is located inside the customer's area.

Ostermeier et al. (2022) considered the real-world limitations and presented an approach to optimize routing with time windows to minimize the total costs of a delivery tour for a given number of available robots. They found the use of trucks and robots for delivery can save costs. These costs are mainly derived from the labor cost of truck drivers and the purchase cost of robots. Chen et al. (2021) discussed the routing problem of scheduling vehicles and delivery robots with time windows. In the description of the problem, some customers can only be served by the vehicles and the rest of the customers can be served by both vehicles and robots. The objective of the problem is to minimize the route time.

## 2.3 Continuous approximation

Solving routing problems requires high computational costs, which limits their real-world applications. However, during the strategic planning phase, such explicit operational details are not essential. For example, deciding whether to locate hubs or the size of the robot fleet can be determined without needing to know which routes they are serving on a particular day. Rather, understanding the generalized trade-offs between different factors is much more crucial. Therefore, the use of closed-form expressions to approximate average or steady-state route performances becomes more relevant, which is the reason why continuous approximation (CA) plays an important role in conducting large-scale policy analysis for strategic planning.

CA uses geometric probabilities of service regions to estimate the expected routing performances. For example, knowing that a vehicle serves ten orders distributed in a uniformly random way across an area, the average distance to serve the ten orders with proper routing could be analytically derived. The derived analytical expression avoids using details of the network (which actual routes are served under a given realization), instead just outputting the steady state performance of the tour generated.

The closed-form expression that approximates the distance of the Travel Salesman Problem (TSP) was first proposed by Beardwood et al. (1959). The formulation is further extended into other forms to address various routing problems. Stein (1978a, 1978b) studied complex pickup and delivery routing considerations (as a Dial a Ride Problem) in which the pickup and drop-off locations are randomly distributed. Daganzo (1984) proposed an effective approximation formulation to estimate the length of a set of routes for a Capacitated Vehicle Routing Problem (CVRP) with random order locations. Subsequent studies based on Daganzo's formulation (Chien, 1992; Kwon et al., 1995; Langevin et al., 1996) have also shown the relative accuracy CA methods can provide. To improve the accuracy of CA methods, additional factors such as the geometry features (Kwon et al., 1995; Robusté et al., 2004; Cavdar and Sokol, 2015) and operation parameters (Diana et al., 2006; Figliozzi, 2008, 2009; Bergmann et al., 2020) have been added to the original CA formulations.

Because of CA models' capability of generating reliable estimation of system-wide performances without operation details and computational complexities, problems focusing on hypothetical logistic system design and evaluation have been proposed based on CA methods. Lemardelé et al. (2021) used CA formulations to calculate the potential operations cost and externalities of deploying drones and autonomous ground vehicles in different cities. Jennings and Figliozzi (2019) studied the potential impact of road autonomous delivery robots by using CA to estimate the performance outcomes under different deployment scenarios. Banerjee et al. (2022) designed a service region partition method based on CA results to enhance the same-day delivery service. Jaller and Pahwa (2020) developed a cost-based analysis tool to evaluate multi-echelon delivery operations involving different vehicle types based on adjusted CA models.

Most CA studies involve evaluating traditional last-mile delivery systems, which consider delivery stops only. However, the application of CA models on more complex delivery operations such as OFD services is lacking. CA models focusing on such pickup and drop-off combined operations have been proposed, but they are not widely applied in analytical models evaluating last-mile delivery systems. Specifically, proper ways of determining the OFD system cost by using the CA model have not been well established. In this study, we propose a model that evaluates the system cost of an OFD service scheme by using an adjusted CA model, which properly addresses the additional distance induced by the added operation of pickup services.

## 3. Methodology

We propose a model that estimates the total cost of an SROFD system. The cost function captures both the fixed cost of fleet purchasing and the SDR operation cost. We extend the Bergmann's model (Bergmann et al., 2020) to specifically estimate the operational costs associated with deploying an SROFD system while considering the constraints of its operations. First, a detailed description of the proposed sidewalk robot delivery system is elaborated. A minimization problem based on the service design optimizing the robot fleet size and the delivery strategies is then presented, from which the optimal values can be analytically derived.

**3.1 Service design**

We consider an SROFD system that provides services in a predetermined area assigned to a depot. The depot serves as a dispatch and charging center for the SDRs. The daily operation is divided into multiple time intervals, and all the orders received in each interval will be delivered by the SDRs in the following interval, as illustrated in Fig. 1. Such an interval-based business model is adapted in multiple OFD platforms such as Instacart and Chowbus. By accumulating the order beforehand, more efficient routes can be pre-organized and reducing detour distance and wandering effect. We formally characterize the system as follows.

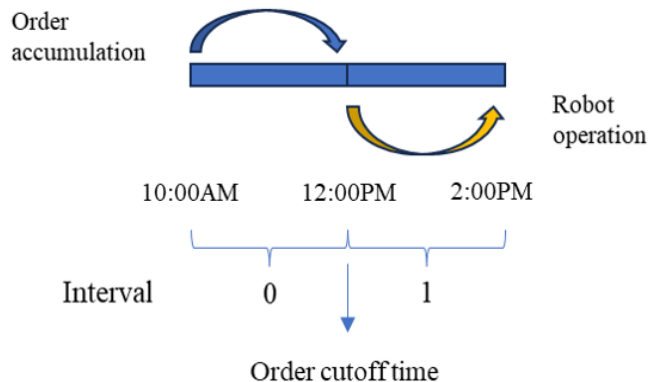

Figure 1. Illustration of a 2-interval service.

- The operational period has a length of $T$ and is divided into $N$ intervals.
- In the first interval '0', orders generated in the two-dimension service region $R$ are received until time $\frac{T}{N}$ and no robot is dispatched for delivery.
- In the following interval $i \in \{1, \ldots, N\}$, SDRs deliver $C_{i-1}$ orders accumulated in interval $i-1$.
- In the final interval $N$, no order is allowed to be received, and only the delivery services are provided to fulfill the $C_{N-1}$ orders.
- The number of orders $C_i$ accumulated during interval $i \in \{0, \ldots, N-1\}$ is assumed to be deterministic for simplicity while the order locations are random. The model can be expanded to incorporate stochastic order receiving process in future studies to obtain operation insights.
- Sidewalk robots start their operations at the start of each interval $i \in \{1, \ldots, N\}$ from their depots. The accumulated orders $C_{i-1}$ are organized into routes for the sidewalk robots to perform the order pickup and drop-off services.
- We assume the system employs $f$ identical robots and the robots traverse the service region $G$ having a constant speed $v_R$.
- During the operation at interval $i$ ($i > 0$), each robot receives battery charging only by the end of the interval and is not available until the start of the next interval $i+1$. Therefore, the range capacity $R$ of each robot should be greater than its traveled distance $\frac{L}{f}$ in each interval.

The following section presents a cost-based minimization problem based on the previously defined service design. It captures both the fixed cost and the operation cost. The robot fleet size during each interval is the decision variable. The cost function is inspired by the one proposed in Jaller and Pahwa (2020). While Jaller and Pahwa capture the fixed and operation costs of a delivery system, they do not capture the operation scenario of an OFD system or the unique characteristics of SDRs. We propose a cost function that addresses the SROFD specified elements including the OFD routing distance and constraints of SDRs.

### 3.2 Cost-based minimization problem

To develop a cost-based model dedicated to an SROFD system, different cost elements need to be properly reflected. Fig. 2 illustrate an SROFD operation in one interval. To complete serving 14 orders in a region, a fleet size with three SDRs is determined. When operating the SDRs, potential maintenance cost needs to be captured. In addition, the energy cost also contributes to the system cost. Both elements can be directly associated with the traveled distance during their operations. Therefore, the required route length to visit all 28 stops in Fig. 2 are needed to calculate the operation cost. The model should also consider the service constraints. The route length should not exceed the SDR's range capacity. Besides, the time spent in service operation and battery charging should also be within the interval's duration.

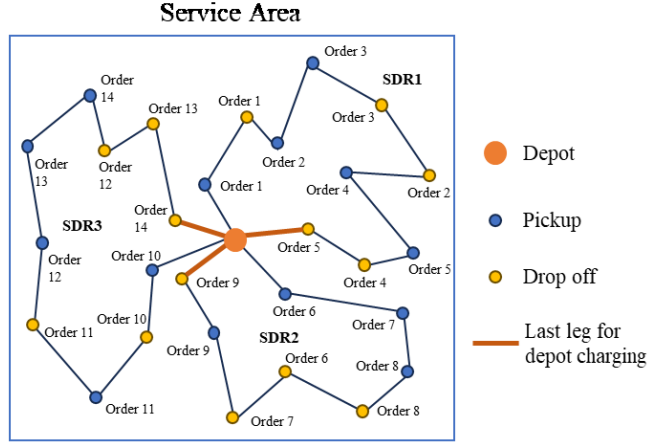

Figure 2. Illustration of an SROFD service in one interval.

We assume the accumulated orders $C$ are uniformly distributed across the service region $G$ and the duration of the interval is $t$. Since the proposed delivery service focuses on food and grocery deliveries, the routes integrate both pickup and drop-off operations. For each route, proper stop sequencing is required to arrange corresponding pickup stops before dropping orders off. A CA model would need to capture the added precedence constraints. One such model was proposed by Bergmann et al. (2020) to capture the additional route distance caused by the pickup integration, which is presented in Eq. (1).

$$L = k^+(l^D + l^P) \qquad (1)$$

$L$ is the average distance of the integrated route, $l^D$ is the pickup only in-region route distance, and $l^P$ is the drop-off only in-region route distance.

According to (Bergmann et al., 2020), the least effective way of routing pickup and drop-off stops is to perform pickup and drop-off services in two separate routes. In other words, the vehicle serves all pickup stops first and subsequently visits all drop-off stops. In such way, the total route length is the longest. The coefficient $k^+$ is a factor that considers the efficiency gain of integrating pickup and drop-off stops into one route. $k^+$ is calculated by using Eq. (2).

$$k^+ = \frac{\sqrt{1+2\beta}}{\sqrt{1-\alpha+\beta}+\sqrt{\alpha+\beta}} + c_1(\sqrt{1-\alpha}+\sqrt{\alpha}-1)\left(\frac{1}{q}-\frac{1}{2}\right) + c_2\frac{\beta}{\sqrt{1+\beta^2}} \qquad (2)$$

where,

$$\alpha = \frac{n_p}{n_p + n_d}, \qquad (3)$$

$$\beta = \frac{n_{sc}}{n_p + n_d}, \qquad (4)$$

$$q = \frac{Q}{max(n_d + n_{sc}, n_p + n_{sc})}. \qquad (5)$$

We closely follow Bergmann et al. (2020) for the notations used in Eq. (2-5). Stop $n_p$ and $n_d$ are the number of pickup and drop-off only stops. They are the "traditional" stops that use the depot as their order drop-off location and order-loading location respectively. $n_{sc}$ represents the number of short-circuiting (SC) orders. Those orders are the ones that need to be first picked up at a stop (e.g., shop/restaurant) and subsequently dropped off at another stop along the route. In our case, the OFD orders are the SC orders, and the SDRs only offer OFD services. Therefore, $n_p = n_d = 0$ and $n_{sc} = C$. We set the number of orders per pickup stop to be $\sigma_p$ and the number of orders per drop off stop to be $\sigma_d$. Assuming that $\sigma_p = \sigma_d = 1$, the SC orders require $C/\sigma_p = C$ pickup stops and $C/\sigma_d = C$ drop off stops. $q$ is the capacity ratio, which calculates the ratio between the robot compartment capacity $Q$ and the maximum of the total loads it needs to handle.

**Proposition 1**. *Given $n_p = n_d = 0$, Eq. (2) becomes a constant.*

**Proof**. When $n_p = n_d = 0$, $\alpha \to 0$, $\beta \to \infty$. This results in Eq. (6).

$$\lim_{\alpha \to 0, \beta \to \infty} k^+ = \frac{\sqrt{2\beta}}{\sqrt{\beta}+\sqrt{\beta}} + c_2 \frac{\beta}{\sqrt{\beta^2}} = \frac{1}{\sqrt{2}} + c_2 \tag{6}$$

This concludes the proof. □

According to Bergmann et al. (2020), $c_2 \approx 0.1042$. Therefore, $k^+ \approx 0.81131$. The $l^D$ and $l^P$ are calculated by using the CA formulation from Beardwood et al. (1959), which is presented in Eqs. (7,8).

$$l^D = k\sqrt{n^D A} \tag{7}$$
$$l^P = k\sqrt{n^P A} \tag{8}$$

$l$ is the route distance of in-area services and $A$ is the size of the service region $G$. $k \approx 0.763$ (Stein 1978a) when using Euclidean metrics and $k \approx 0.97$ (Jaillet 1988) when using Manhattan metrics. $n$ is the number of stops, and $\delta$ is the stop density in region $R$. The total number of pickup stops $n^P$ equals to $n_p + n_{sc}$, and the total number of drop off stops $n^D$ equals to $n_d + n_{sc}$. With $n_p = n_d = 0$, $n^D = n^P = n_{sc} = C$. By applying Eqs. (6-8), Eq. (1) can be re-written as Eq. (9).

$$L = 1.62262(k\sqrt{AC}) \tag{9}$$

We formulate a total cost $TC$ minimization problem (Eq. 10) with predetermined service region $G$ and depot $O$. It considers both the fixed cost and the operation cost in a single interval. The first term of Eq. (10) is the fixed cost $P$ including the robot purchasing cost. The second term captures the operation cost including a per-mile maintenance cost and an electricity charging cost. The per-mile maintenance cost is calculated by multiplying the tour length $L$ with the maintenance cost per mile $\pi_d$. The electricity charging cost is calculated by multiplying the tour length $L$ with the energy consumption per mile $r_c$ and the per unit energy cost $\pi_c$. The decision variable is the dispatched robot fleet size $f$. We do not consider the setup cost of a depot in the initial calculation. However, the marginal cost of opening additional depots can indicate what would be the acceptable cost of adding an extra depot.

$$\min_{f} \quad TC = Pf + (L\pi_d + Lr_c\pi_c) \tag{10}$$

$$\text{subject to} \quad \frac{\varrho L}{fv_R} + \frac{C}{f}(\tau_p + \tau_d) + \varphi\frac{\varrho Lr_c}{f\mu_c} \leq t \tag{11}$$

$$\varrho\frac{L}{f} \leq R \tag{12}$$

Eq. (11) ensures the time requirement of serving $C$ orders in one interval. Since $L$ could only provide the average tour length, the coefficient $\varrho > 1$ captures the routes having a longer-than-average distance. According to the fitting result conducted in (Figliozzi, 2008), the length of a route is generally shorter than 1.25 times the average route length estimated by a CA model. Therefore, we set $\varrho = 1.25$ to account for extreme cases. The first term calculates the average travel time with the added buffer $\varrho = 1.25$, and the second term considers the service time required to complete both pickup and drop-off services. The average route distance $L$ is calculated using Eq. (9). The third term is the battery recharge time, which is calculated by using total energy consumption divided by the recharge rate $\mu_c$. We add the $\varrho = 1.25$ to also ensure that all robots can finish their charging session before the end of an interval. When the battery is close to full, the charging power would drop significantly for a common Lithium-ion battery (Lee et al., 2008). Therefore, the extra coefficient $\varphi >= 1$ is used to count the additional charging time when the battery is close to full. Eq. (12) ensures the tour length is within the robot's range capacity.

The optimal fleet size $f^*$ can be directly derived by combining Eqs. (10-12) by formulating the Lagrangian. We formally present the process in the following proposition.

**Proposition 2**. *Given Eqs. (10-12), the optimal fleet size $f^*$ and optimal cost $TC^*$ are shown in Eqs. (13) and (14).*

$$f^* = \frac{1}{t}\left(\frac{\varrho L}{v_R} + C(\tau_p + \tau_d) + \varphi\frac{\varrho Lr_c}{\mu_c}\right), \tag{13}$$

$$TC^* = \frac{P}{t}\left(\frac{\varrho L}{v_R} + C(\tau_p + \tau_d) + \varphi\frac{\varrho Lr_c}{\mu_c}\right) + (L\pi_d + Lr_c\pi_c) \tag{14}$$

***Proof***. By applying Lagrangian multipliers $\lambda_1$ and $\lambda_2$ to Eq. (11) and Eq. (12) respectively and adding them back to Eq. (10), we have Eq. (15).

$$\min_{f} TC$$
$$= Pf + (L\pi_d + Lr_c\pi_c) + \lambda_1(\frac{1}{t}\left(\frac{\varrho L}{v_R} + C(\tau_p + \tau_d) + \varphi\frac{\varrho Lr_c}{\mu_c} - f\right) + \lambda_2\left(\varrho\frac{L}{R} - f\right) \tag{15}$$

After taking the derivatives of Eq. (15) corresponding to $f$, $\lambda_1$, and $\lambda_2$, and then setting them to equal to zero, we get the first order conditions shown in Eqs. (16-18).

$$\lambda_1^* + \lambda_2^* = P, \tag{16}$$

$$\lambda_1^*\left(f^* - \frac{1}{t}\left(\frac{\varrho L}{v_R} + C(\tau_p + \tau_d) + \varphi\frac{\varrho Lr_c}{\mu_c}\right)\right) = 0, \tag{17}$$

$$\lambda_2^* \left( f^* - \varrho \frac{L}{R} \right) = 0. \tag{18}$$

We assume that the route distance will always be shorter than the SDR's range in a compact area, which is further illustrated in the case study. In such case, Eq. (12) becomes non-binding, making $\lambda_2 = 0$. Therefore, $\lambda_1 = P$, and Eq. (13) is the optimized fleet size. Plugging this fleet size back into Eq. (10) results in Eq. (14). This concludes the proof. □

Based on Eq. (14), the relationship between the system cost and the different variables involved in the model can be evaluated. The following section discusses the impact of variables on cost in detail.

### 3.3 Trade-off analysis

In this section, we use the derived formulation to analyze how different variables would interact with the system cost. It offers an in-depth look at the SROFD system, shedding light on future applications when the decision-makers are facing different operation conditions and strategies. The average cost per order indicates the level of delivery fee each customer would pay when placing an OFD order. Therefore, the following analysis is conducted based on average cost $AC$. By dividing Eq. (14) with number of orders $C$, the optimal average cost $AC^*$ is shown in Eq. (19).

$$AC^* = 1.62262 \left( k\sqrt{A/C} \right) \left[ \frac{P}{t} \left( \frac{\varrho}{v_R} + \varphi \frac{\varrho r_c}{\mu_c} \right) + (\pi_d + r_c \pi_c) \right] + \frac{P}{t} (\tau_p + \tau_d). \tag{19}$$

**Corollary 1**. *The SROFD system exhibits economies of scale.*

*Proof.* By taking the first derivative of Eq. (19) on $C$, we get Eq. (20).

$$\frac{d}{dC} AC^* = -0.81131 \frac{k\sqrt{A}}{C\sqrt{C}} \left[ \frac{P}{t} \left( \frac{\varrho}{v_R} + \varphi \frac{\varrho r_c}{\mu_c} \right) + (\pi_d + r_c \pi_c) \right]. \tag{20}$$

As shown in Eq. (20), $AC^*$ decreases when the demand increases. This concludes the proof. □
When $C$ increases while other variables are being fixed, the downward slope of $AC$ becomes smaller, showing a decreasing marginal cost reduction. For example, the slope when $C = 100$ is 2.83 times the slope when $C = 200$, and the value becomes 31.62 when compared to the slope when $C = 1000$. Therefore, it is crucial for the operator to increase its market size in its early phase to significantly reduce the average cost per order, making its service cheaper and more competitive.

When the demand is already at a higher level, a more effective way of lowering the average cost is to reduce the service time at each stop. As shown in Eq. (20), the relationship between $AC$ and stop service time is linear. When $C$ is at a higher level, the proportion of the value from the first term decreases in $AC$. Therefore, a unit decrease of the second term caused by the reduction of $\tau_p$ and $\tau_d$ would play a more significant role in the equation.

We also evaluate the impact of the duration of an interval. We assume that during an interval, the order receiving rate per unit time $\delta$ is a constant. Therefore, Eq. (20) can be written as Eq. (21).

$$AC^* = 1.62262 \left(k\sqrt{A/(\delta t)}\right) \left[\frac{P}{t}\left(\frac{\varrho}{v_R} + \varphi \frac{\varrho r_c}{\mu_c}\right) + (\pi_d + r_c\pi_c)\right] + \frac{P}{t}(\tau_p + \tau_d). \tag{21}$$

By increasing the duration of an interval, a more significant decrease in $AC^*$ can be observed compared to the cost reduction caused by $C$. Instead of only affecting the first term in $AC^*$ formulation, $t$ also affects the second term. Therefore, when the operator is struggling with increasing the demand of orders, a longer service interval can be a viable option to reduce the service cost if it does not impact demand too negatively. However, longer interval means longer order wait time, which could in turn make the service less attractive. Such demand elasticity patterns are not explicitly incorporated in the model and can be further investigated in future studies.

The battery charging speed also contributes to the level of $AC^*$. By increasing the charging rate, lower $AC^*$ can be achieved. The first order derivative of Eq. (19) on $\mu_c$ is shown in Eq. (22).

$$\frac{d}{d\mu_c} AC^* = -1.62262\left(k\sqrt{A/C}\right)\left(\frac{P}{t}\varphi \frac{\varrho r_c}{\mu_c^2}\right). \tag{22}$$

Again, a diminishing effect of cost reduction can be observed when the charging speed is higher. Therefore, it is more cost-effect to use a more powerful charging hardware when the base charging rate is low. In addition, Eq. (22) becomes smaller when $C$ is at a lower level.

Eq. (13) is based on the assumption of route distance per SDR being lower than the range capacity. The tour length per robot $L_{avg}$ can be calculated by dividing Eq. (9) by Eq. (13), which is presented in Eq. (23).

$$L_{avg} = t\left(\frac{\varrho}{v_R} + \sqrt{C}/(1.62262k\sqrt{A})(\tau_p + \tau_d) + \varphi \frac{\varrho r_f}{\mu_c}\right)^{-1}. \tag{23}$$

$L_{avg}$ decreases when $C$ increases. Therefore, as long as the range capacity is higher than the route length when the demand is low, Eq. (13) always holds true when the demand is higher. In the case study, we quantify the required range capacity to further confirm this assumption.

In the current service design, the charging operation is included in each interval. However, higher cost savings can be achieved by moving the charging session out of the service interval. In real-world application, this could mean that the depot closing time is after the end of the delivery interval plus the robot charging time. In such case, the equation for calculating $AC^*$ becomes Eq. (24).

$$AC^* = 1.62262 \left(k\sqrt{A/(\delta t)}\right) \left[\frac{P\varrho}{tv_R} + (\pi_d + r_c\pi_c)\right] + \frac{P}{t}(\tau_p + \tau_d). \tag{24}$$

It means the cost saving would be higher compared to only increasing the charging speed by completely dropping the charging session out of the cost function.

Since the system utilizes SDR for delivery, the level of service (LOS) of the sidewalk has a direct impact on the robot travel speed, which inadvertently affects the en-route travel time. According to a preliminary study conducted by Li et al. (2023), the delivery robot speed would drop by 50% when the pedestrian LOS drops from A to D. Because $v_R$ only affects a portion of the cost calculated in the first term in Eq. (19), the average cost increase would be smaller than the 50% speed reduction. However, the cost would still significantly increase when the SDRs are operating on crowded sidewalks. In the case study, we quantify such cost increases for better illustration.

## 4. Case study

### 4.1 Data and problem setting
After investigating the impact of variables on the average cost, we apply the model in a real-world scenario. We use three Neighborhood Tabulation Areas (NTAs) in Manhattan, NY (Fig. 3) to conduct our case study, which covers an area of 3.51 square miles. Synthetic OFD orders are generated based on the number of households in the neighborhoods and estimated daily orders per household. In total, 86745 households reside in the area (NYC Planning, 2023). In the US, the average yearly spending of an OFD user is $1843.72, and the average cost per order is $33.89 (Circuit, 2023). Therefore, an average of 0.15 orders would be placed by a user per day. In 2022, the market penetration of OFD in the US was 43.9 percent, (Statista, 2023). We assume that each user represents one household, and the national average applies to the residents in the study area. Therefore, the whole area is estimated to generate 5712 daily OFD orders.

We assume that each robot costs $5,000, which is at the upper bound of what a UC Berkeley report revealed (Berkeley, 2023). Therefore, the average daily depreciation cost would be $6.85 assuming a 2-year lifespan and 0 salvage value. We use the average maintenance cost of an EV provided by the federal Office of Energy Efficiency and Renewable Energy as the standard maintenance cost of an SDR, which is $0.06 per mile (EERE, 2023). We use the average electricity cost in NYC to calculate the energy bill, which is 0.17/kWh according to NYSERDA (2023). The value of $\varphi$ in Eq. (11) is assumed to be 1.2 and a charging rate of 400 W/h is applied. In terms of robot specs, we use some of the Starship robot's configurations (Starship, 2023). The maximum running speed is 4 miles/hr, and the average power consumption is estimated to be 28.7 Wh/mile. We assume the battery capacity is 200 Wh, which is equivalent to a range capacity of 6.97 miles. Manhattan metrics is applied to the model. Therefore, we set $k \approx 0.97$ (Jaillet, 1988).

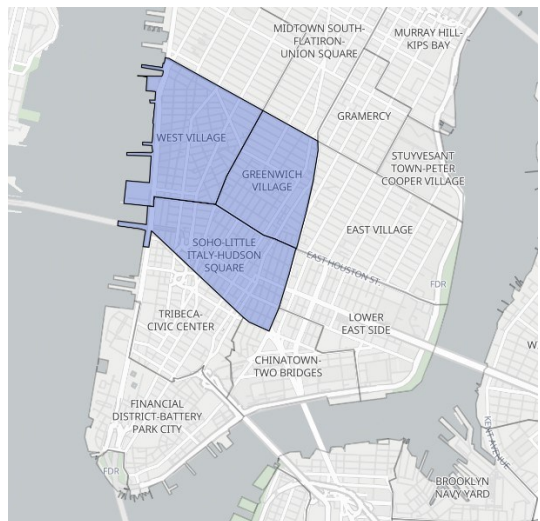

Figure 3. Studied neighborhood in Manhattan.

We assume that one depot at the center of the area provides the SROFD service. The service is deployed during the lunch peak hour. The system starts to receive orders from 10 AM to 12 PM, and the SDRs start to deliver the received orders from 12 PM to 2 PM. Assuming 5% of the total orders are collected during the 2-hour interval, this equals 285 orders. Each pickup stop requires a 3-minute service time, and a 2-minute service is applied to each drop-off stop. In practice, drop-offs for OFDs might require longer for high-rise apartment buildings, but for delivery robots, we can assume that drop-offs will only be done at the building lobby.

### 4.2 Fleet deployment results
By applying the values to Eq. (13) and Eq. (14), 18.41 robots are needed to fulfill the service requirement. The total cost is $128.08 for the interval, and the average cost per order is $0.45. The average route length per robot is 1.68 miles, which is significantly lower than the SDR's range capacity. Even if we multiply $\rho$ with the average length, an "extra-long" route with 2.1 miles is still much smaller than the range capacity. According to Eq. (23), the range capacity can always cover the route length when the demand is higher.

### 4.3 Sensitivity to demand and service times
Fig. 4 shows the change of average delivery cost by varying the share of total demand at the studied interval, along with a pickup service time change. When the pickup time remains 3-minute long, the average cost becomes $0.38 per order when 15% of the daily order is delivered during the interval. With 2-minute pickup time, the average cost becomes $0.39 per order when the demand share is 5%, and it decreases to $0.32 per order when the share increases to 15%. The cost saving is 12.74% by reducing the pickup service time by 1 minute when the demand share is 5%, and the number becomes 15.05% when the demand share is 15%. Therefore, when the demand level is higher, it is essential to improve the operation efficiency by reducing the service time at each stop.

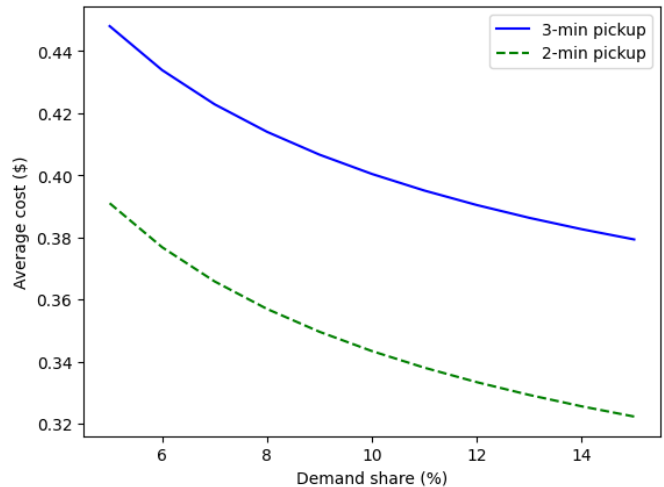

Figure 4. Impact of demand and pickup service time on average cost per order.

### 4.4 Sensitivity to charging rate
Fig. 5 illustrates the average cost saving when having various levels of charging rates under different demand levels. All other variables are kept unchanged. By increasing the recharge rate from 400 W/h to 2 kW/h, the cost saving is 7.11% when the demand share is 5%. Such saving shrinks to 4.85% when the demand share increases to 15%. If the charging session is moved out of the service interval, the cost saving would be 8.90% when the demand share is 5%. It reduces to 6.07% when the demand share increases to 15%. In comparison, the cost saving is slightly higher by excluding the charging session in the service interval.

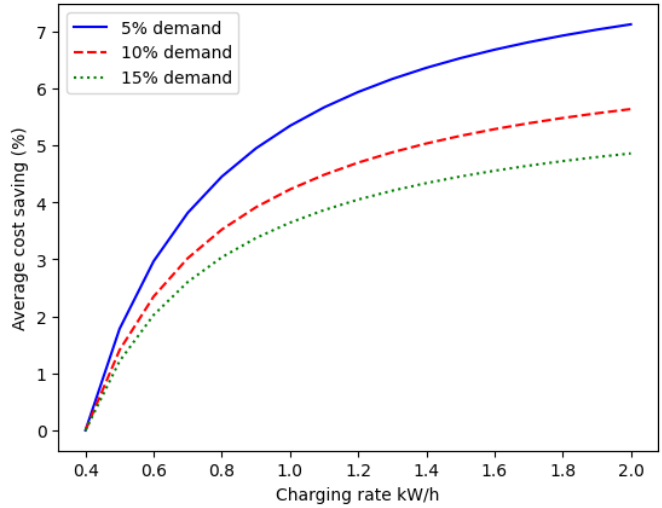

Figure 5. Impact of battery charging rate on (a) daily average cost and (b) average cost savings under various demand levels.

### 4.5 Sensitivity to pedestrian densities
As mentioned in the previous section, the crowding condition on the sidewalk could have a significant impact on the cost. We set the 4 miles/hr speed as the SDR LOS A speed. Therefore, the LOS D speed becomes 2 miles/hr by applying the 50% speed reduction coefficient. With a 5% demand share, the cost is 25.84% higher when running on LOS D sidewalks. The number drops to 17.62% when the demand share is 15%. Therefore, it is favorable to deploy an SROFD system in a busy area when the demand is high, and a sudden surge in pedestrian density (e.g., street fair) would significantly impact the service cost.

### 4.6 Depot design strategy analysis
To further investigate how the change of service strategy would impact the average cost, we study a scenario where three depots located at the centers of the neighborhoods provide the service in their local areas. Table 1 is the summary of each neighborhood's geographic features and the corresponding model results with 5% demand share. The system-wise fleet size and total cost only experience slight change, and the marginal gain of adding a new depot in a relatively compact area is therefore nearly negligible. In such case, only using one depot near the center of the whole area to conduct the service is a better solution.

Table 1. Results of using three neighborhood depots to provide OFD services

| Name | Area (sq. mi.) | Orders | Optimal fleet size | Total cost ($) |
|---|---|---|---|---|
| SoHo-Little Italy-Hudson Square | 0.46 | 74 | 5.01 | 34.93 |
| Greenwich Village | 0.38 | 99 | 6.15 | 42.77 |
| West Village | 0.51 | 112 | 7.17 | 49.86 |
| Sum | 1.35 | 286 | 18.33 | 127.56 |

It is unrealistic to exclude the cost of establishing a new depot. After searching online, a typical storefront commercial space cost $200 sq ft per year in the study area. We assume that a 500 sq ft space is rented for the SROFD operation, the yearly rent cost is thus $100,000, which equals to $273.97 daily cost. By adding the depot rent, the total cost becomes $329.83 and the average cost is therefore $1.41 per order. When the demand share increases to 15%, the average cost would drop to $0.70. In both cases, the average costs are significantly lower than the current OFD delivery fee, showing the huge potential of deploying SDR to provide OFD services in the future.

## 5. Conclusion

The OFD industry has grown drastically in recent years. With higher OFD demand, the challenge of providing last-mile services has become more pronounced. Various optimization models have been developed to increase the overall system efficiency, while other means of delivery methods have been explored. Among them, using SDRs to conduct the last-mile service has gathered increasing attention, and multiple pilot programs have been conducted to evaluate its operational feasibility. However, there has been limited work that systematically evaluates the cost of deploying SDRs to provide OFD services and how the system could be impacted by different variables. In this study, we propose a model that estimates the total cost of an SROFD system by optimizing robot fleet size within a given service area and order demand. The depot is assumed to locate near the center of the service area. The objective function and constraints are formulated based on a CA model that addresses the additional distance caused by integrating pickup and drop-off stops in one delivery route. A closed-form expression is derived to directly calculate the optimal fleet size and system cost. By applying the model with given inputs of variables, decision makers can easily estimate the impacts of different policies and strategies on potential SROFD applications.

We inspect how different variables could impact the system cost based on the derived formulations. The model is then applied to an area containing three Manhattan neighborhoods to estimate the real-world application of deploying an SROFD system. The key findings are presented as follows:

- A diminishing effect of the economies of scale occurs when the demand level increases. It is thus important to expand the market share aggressively to reduce the system cost at an early stage.
- The time spent at each service stop has a significant impact on the total cost. The higher the demand level is, the higher the cost saving is observed by shortening the service time at stops.
- Increasing the battery charging rate has a limited effect on reducing the average cost. The cost saving from increasing the battery charging rate is more significant at lower demand levels.
- Moving the charging session out of the service interval has a greater impact on cost reduction compared to increasing the charging speed. However, the improvement would also be limited.
- The sidewalk LOS can significantly affect the cost. When the sidewalk LOS drops from A to D, the cost would reduce by 10% to 30% depending on the demand level. Therefore, the deployment of SROFD service in busy areas (e.g., CBD) requires more deliberation.
- When the service area is compact, no added benefit can be observed by opening more local depots. This observation may be tempered by larger neighborhood aggregations in which tour length trade-offs become more significant or where neighborhoods exhibit greater heterogeneity.

- By applying the SROFD system in a compact area such as the neighborhoods in Manhattan, the average cost per order is significantly lower than the delivery charge requested by current OFD platforms at different demand levels even considering the depot rental cost. Therefore, the deployment of SROFD systems could bring significant benefits to the OFD industry.

The developed model can be further improved when more OFD operation details are available. The model can be used to evaluate the impact of different service elements and deployment strategies. The model is developed based on the assumption of deterministic order demands (with stochastic pickup and drop-off locations), which only provides a high-level estimation of the system cost. The model can be extended to incorporate stochastic demand in future studies. An interval-based service design is assumed in this study, and we only focus on a single-interval scenario. Multi-interval scenarios with dynamic dependencies can be investigated in future studies, which could provide a better estimation of the system cost. Further studies are needed to develop more complex analytical models to better evaluate the system using the on-demand strategy.